\documentclass[aps,prx,superscriptaddress,twocolumn]{revtex4-2}
 \usepackage{graphicx,color}
 \usepackage{verbatim}
 \usepackage{amssymb}   
 \usepackage{amsmath}
 \usepackage{amsfonts}
 \usepackage{mathdots}
 \usepackage{hyperref}
 \usepackage{epsfig}
 \usepackage{hyperref}
 \usepackage{xcolor}
 \definecolor{myblue}{RGB}{46, 48,146}
 \usepackage{lineno}
 \hypersetup{colorlinks=true,linkcolor=myblue,citecolor=myblue,urlcolor=myblue, linktocpage}
 \usepackage{braket}
 \usepackage{bm}
 \usepackage{multirow}
 \usepackage{comment}

\begin{document}
 	\title{ Nonreciprocal Nonlinear Responses   in   Moving 
 Charge Density Waves }
 	
 	\author{Ying-Ming Xie}\thanks{email: yingming.xie@riken.jp}
    \affiliation{RIKEN Center for Emergent Matter Science (CEMS), Wako, Saitama 351-0198, Japan}
    \author{Hiroki Isobe}
    \affiliation{RIKEN Center for Emergent Matter Science (CEMS), Wako, Saitama 351-0198, Japan}
    \author{Naoto Nagaosa}\thanks{email: nnagaosa.naoto@googlemail.com}
   \affiliation{RIKEN Center for Emergent Matter Science (CEMS), Wako, Saitama 351-0198, Japan}
   \affiliation{Fundamental Quantum Science Program, TRIP Headquarters, RIKEN, Wako 351-0198, Japan}

\begin{abstract}
{\bf Abstract} The incommensurate charge density wave states (CDWs) can exhibit steady motion in the flow limit after depinning, behaving as a nonequilibrium system with time-dependent states. Since the moving CDW, like an electric current, breaks both time-reversal and inversion symmetries, one may speculate the emergence of nonreciprocal nonlinear responses from such motion. However, the moving CDW order parameter is intrinsically time-dependent in the lab frame, and it is known to be challenging to evaluate the responses of such a time-varying system.      In this work, following the principle of Galilean relativity, we resolve this time-dependent hard problem in the lab frame by mapping the system to the comoving frame with static CDW states through the Galilean transformation. We explicitly show that the nonreciprocal nonlinear responses would be generated by the movement of CDW states through violating Galilean relativity. 


 	\end{abstract}
 	
 	\date{\today}
 	
 	\maketitle


\noindent {\bf Introduction} Nonreciprocal nonlinear responses can manifest when inversion symmetry (and time-reversal symmetry in many cases)  is broken \cite{Tokura2018, Nagaosa2024}. These phenomena have garnered significant theoretical and experimental attention in recent years, notably in phenomena such as nonlinear Hall effects \cite{Liang2015,Du2021} and nonreciprocal superconducting effects \cite{Ryohei2017, Ando2020, Nadeem2023,Nagaosa2024}. However, while much focus has been near the equilibrium states, nonreciprocal nonlinear responses in nonequilibrium systems with time-dependent states remain largely unexplored. Theoretical investigation of such responses in nonequilibrium states is challenging due to the inherent time dependence and dynamic nature of these systems.

In condensed matter physics, current-driven systems can exhibit nonequilibrium behavior and intrinsic time-dependence. For example, certain symmetry-breaking orders would undergo motion once they surpass impurity-pinning effects under an electric field beyond the threshold. A notable example is the incommensurate charge density wave (CDW), where intriguing dynamical properties emerge upon depinning  \cite{LEE1974,Patrick_Lee1,Patrick_Lee2,Sneddon1982, Gruner1988}. In the limit of large current, these CDW states flow steadily, representing a nonequilibrium steady state.

\begin{figure}
	\centering
	\includegraphics[width=1\linewidth]{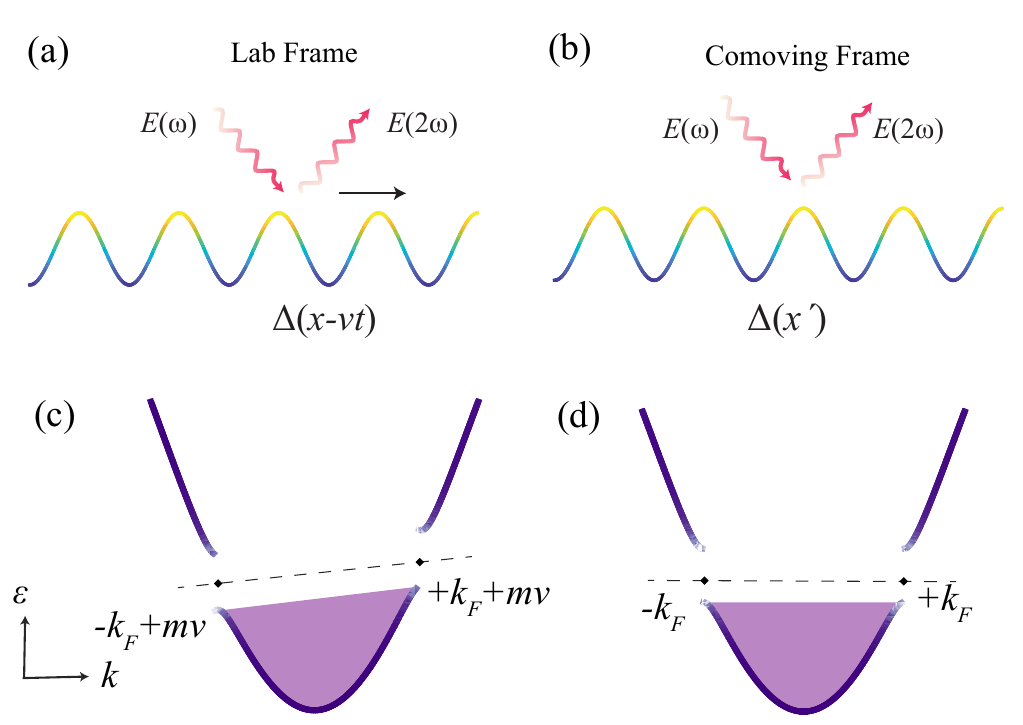}
	\caption{ Illustration of the CDW states in the comoving frame and the lab frame. (a) and (b) schematically show the CDW order parameter in the comoving and lab frame, respectively. The black arrow in (a) represents the CDW motion. The possible nonreciprocal nonlinear responses induced by the moving CDW states are highlighted as the second harmonic generation. (c) shows the moving CDW states in the lab frame, where the Fermi momentum is shifted to $\pm k_F+ mv$ and the CDW gap opens near these shifted Fermi momenta, while (d) shows the CDW states in the comoving frame, where the CDW gap opens at $\pm k_F$. Note that (c)  cannot be simply regarded as the real energy dispersion due to the generic time-dependent feature in the lab frame. } 
	\label{fig:fig1}
\end{figure}

A CDW motion breaks both inversion and time-reversal symmetries (see illustrations in Figs.~\ref{fig:fig1} (a) and (c)), as implied in the seminar work by Allender, Bray, and Bardeen \cite{Bardeen1974}.    One may naturally ask whether there are any finite nonreciprocal nonlinear responses induced by the current-driven motion  [Fig.~\ref{fig:fig1}(a)].  However, directly addressing this problem is challenging due to the intrinsic time dependence of a moving CDW in the lab frame.

On the other hand, in classical physics, we know that the principle of relativity is often powerful in simplifying a problem by changing inertial frames.  Following this principle,  the moving CDW problem can be reconsidered in the comoving frame, where the CDW state becomes static.  In contrast to the lab frame, the asymmetry induced by the external current appears to vanish in the comoving frame as illustrated in Figs.~\ref{fig:fig1} (b) and (d).  As a result, one may expect that nonreciprocal nonlinear responses vanish in the comoving frame. Looking at the nonreciprocal nonlinear
responses for moving states within both the lab and
comoving frame is indeed puzzling. 

Motivated by exploring nonreciprocal nonlinear responses in nonequilibrium systems with time-dependent state and the aforementioned puzzle, we explicitly study nonreciprocal nonlinear responses in moving CDW states in this work. We begin with a simple one-dimensional (1D) CDW model, where the CDW order parameter moves spatially at a constant velocity. Using the field theory approach, we identify that the moving CDW state can be mapped to a static one through the Galilean transformation.  The transformation is identical to the change of inertial frames from the lab frame to the comoving frame with the CDW.  We also argue that the Galilean transformation dictates the invariance of the optical conductivity in the lab and comoving frames, facilitating the straightforward solution of optical responses in the comoving frame.  Based on this understanding, we show that the nonreciprocal nonlinear responses are absent with Galilean relativity, where the single-particle dispersion is simply quadratic and thus Galilean invariant.  By introducing a quartic term to violate Galilean relativity, we explicitly show that nonreciprocal nonlinear responses would appear. 

\noindent {\bf Results}

\noindent {\bf A moving CDW model and Galilean transformation.} Let us first illustrate how we can map a time-dependent problem in the lab frame onto a time-independent problem in the comoving using the principle of relativity.    We start with the simplest 1D continuum model with a moving CDW, which has been widely used to study the dynamics of a CDW \cite{Gruner1988,Bardeen1974,gruner2018density}:
\begin{equation}
	\mathcal{L}=i\psi^{\dagger}\partial_t\psi-\frac{1}{2m}\partial_x\psi^{\dagger}\partial_x\psi-\psi^{\dagger}V(x-vt)\psi, \label{eq_lab}
\end{equation}
where $\mathcal{L}$ is the Lagrangian,   $\partial_x$ denotes $\frac{\partial}{\partial x}$, $\psi(x,t)$ is a field operator for electrons, $m$ is an effective mass, and  we set $\hbar=1$ in the main text. The CDW order parameter is 
\begin{equation}
V(x-vt)=2\Delta \cos(2k_F(x-vt))
\end{equation}
with $v$ the drift velocity of the CDW. The wave vector of the CDW is denoted as $2k_F$, where $k_F=\sqrt{2m\epsilon_F}$ is the Fermi wave vector at the Fermi energy $\epsilon_{F}$.  The electronic energy dispersion acquires an energy gap $\Delta$ at the Fermi level, corresponding to the CDW order parameter. 
Such CDW states emerge in quasi-1D systems \cite{Gruner1988}. The Lagrangian density $\mathcal{L}$ describes the dynamics of quasi-1D CDW states driven by an electric current in the flow region \cite{Gruner1988,Bardeen1974,gruner2018density}. We shall assume disorder effects are negligible in the flow region.  Note that due to the lack of dissipations induced by the disorder effects, a finite electric field infinitely accelerates electrons.  Instead, what drives the moving CDW state is the DC electric current other than 
  the electric field. In this case, the electric current arises from the time-varying of the CDW phase \cite{Gruner1988}.

To eliminate the time dependence of the CDW order parameter,  we introduce the comoving frame by performing the Galilean transformation on the coordinates  $x'=x-vt, t'=t$, $\partial_t=\partial_t'-v\partial_{x'}$, $\partial_x=\partial_{x'}$. The field operator transforms as $\psi=e^{i \eta(x,t)}\psi'$ with the phase factor $\eta(x,t)=mvx-\frac{1}{2}mv^2 t$; see Method section  for details.  After this transformation,  the Lagrangian in the moving frame reads
\begin{equation}
\mathcal{L}'= i \psi'^{\dagger}\partial_{t'}\psi'-\frac{1}{2m} \partial_{x'}\psi'^{\dagger}\partial_{x'}\psi'-\psi'^{\dagger}V(x')\psi' \label{eq_move}
\end{equation}
with $V(x')=2\Delta \cos(2k_Fx')$. Comparing Eq.~\eqref{eq_lab} with Eq.~\eqref{eq_move}, we can see that the exact Galilean invariance does not exist due to the CDW order parameter. However, the problem in the lab frame with the moving potential $V(x-vt)$ is mapped to the one in the comoving frame with the static potential $V(x')$. As there are no extra new terms after Galilean transformation,  we would call the system exhibits \textit{Galilean relativity} in this case. Because of this Galilean relativity, the nonlinear responses would not appear since the energy dispersion is simply quadratic in the comoving frame without time dependence.  

In general, there is no exact Galilean relativity in solids. In this work, we invoke a violation of Galilean relativity by introducing a quartic term:
\begin{equation}
\delta \mathcal{L}=-\lambda \partial_x^2\psi^{\dagger}\partial_x^2\psi,
\end{equation}
which is allowed by any symmetries.
In this case, the total Lagrangian is $\mathcal{L}_t=\mathcal{L}+\delta \mathcal{L}$. After the Galilean transformation, it is straightforward to show the Lagrangian in the comoving frame becomes
$\mathcal{L}'_t=\mathcal{L}'+\delta \mathcal{L}'$, where $\delta \mathcal{L}'$ represents the quartic term in the comoving frame:  
\begin{eqnarray}
\delta \mathcal{L}'&&=-\lambda \psi'^{\dagger}\partial_{x'}^4\psi'-\lambda m^4v^4\psi'^{\dagger} \psi'+4i \lambda m^3v^3\psi'^{\dagger}\partial_{x'}\psi'\nonumber\\
 &&+6\lambda m^2v^2\psi'^{\dagger}\partial_{x'}^2\psi'-4i\lambda mv\psi'^{\dagger}\partial_{x'}^3\psi'.\label{main_5}
\end{eqnarray}
It can be seen that when we consider a more general dispersion, beyond the simplistic quadratic band, residual terms emerge in the Lagrangian, which cannot be eliminated in the comoving frame. It is also interesting to note that the terms that are odd in momentum in Eq.~\eqref{main_5} break inversion and time-reversal symmetries, which is expected for moving CDWs but forbidden by Galilean relativity previously. As we will see later, the cubic-momentum term, i.e., the last term in Eq.~\eqref{main_5},  results in nonreciprocal nonlinear responses.

It is worth noting that we can map the Lagrangian to be time-independent for the moving CDW model even with the quartic term via Galilean transformation. This allows us to evaluate some interesting effects that previously were hard to demonstrate in a time-dependent system. To be specific, we shall focus on the nonreciprocal nonlinear responses next.   
\vspace{1\baselineskip}

\noindent {\bf The invariance of finite frequency responses under Galilean transformation.} We next show that the nonreciprocal nonlinear responses can be equivalently calculated in the comoving frame. The conductivity of nonlinear responses is defined by expanding the current density in the powers of external fields. For example, the $n$-th order  harmonic generation in the lab frame is obtained from  $J_x(n\omega)=\sigma_{xx}(n\omega) E^{n}_x(\omega)$, where $n$ is a positive integer and $\omega$ is the incident photon energy $(\hbar = 1)$, $E_x$ denotes the electric field along $x$-direction. Note that we have simplified subscript in the conductivity; for example, the second-order conductivity tensor $\sigma_{xxx}(2\omega)$ would be simply labeled with $\sigma_{xx}(2\omega)$ or $\sigma(2\omega)$ below.  

Now we show how the optical conductivity transforms upon a Galilean transformation by using the Galilean transformation relation of the electric field and current. According to the minimal coupling principle,  an electromagnetic field would appear in the Lagrangian by replacing the $\partial_i\psi$ in Eq.~\eqref{eq_lab} as $(\partial_i+ieA_i)\psi$,  $\partial_i\psi^{\dagger}$ as $(\partial_i-ieA_i)\psi^{\dagger}$ \cite{Son2006}. Then the current-density operator is $j_x=\frac{\partial \mathcal{L}_t}{\partial A_x}|_{A_x=0}$.  Upon the Galilean transformation $\psi=e^{i\eta(x,t)}\psi'$, we find that the current-density operators in the lab and  comoving frames satisfy
\begin{equation}
\hat{j}'_x=\hat{j}_x+ev\psi'^{\dagger}\psi', \label{Eq:main6}
\end{equation}
where $\hat{j}_x'$ is the current operator deduced from the Lagrangian $\mathcal{L}'_t$.  See the explicit derivations in 
 Supplementary Note 2. Note that Eq.~\eqref{Eq:main6} holds even with the quartic term.  Sandwiching  the current operator with the ground state, we obtain the well-known Galilean transformation for the current
\begin{equation}
J'_x(t)=J_x(t)+ev N, \label{current}
\end{equation}
where $t'=t$ is used and the total electron number $N=\int dx' \braket{\psi'^{\dagger}\psi'}$ , where $\braket{}$ denotes the average over the ground states. 
The form of Eq.~\eqref{current} is consistent with the one obtained from the relativity theory in classical electrodynamics (Supplementary Note 1).  Moreover, as shown in Supplementary Note 2B, we argue that Eq.~\eqref{current} holds for a generic energy dispersion by performing Galilean transformation in the momentum space, where the group velocity of each electron is uniformly shifted by $v$. 
Operating $\int dt e^{-in\omega t}$  on both sides, we find  $J'_x(n\omega)=J_x(n\omega)$ for $\omega \neq 0$. Here we consider a finite frequency response because the DC conductivity with $\omega=0$ is divergent without dissipations. 
Note that the current $evN$ in Eq.~\eqref{current} does not contribute to the finite frequency response as $N$ is time-independent due to the conservation of total electron number in the system.  
On the other hand, the electric fields of light in the two frames are the same $\bm{E}'=\bm{E}$ when there are no external magnetic fields \cite{jackson1999classical} and Supplementary Note 1, resulting in $E_x(n\omega)= E_x'(n\omega)$. Using $\sigma_{xx}'(n\omega)=J_x'(n\omega)/E'^{n}_x(\omega)$, we find that the optical conductivity in the lab frame and comoving frame are  equal:
\begin{equation}
\sigma_{xx}(n\omega)=\sigma'_{xx}(n\omega).
\end{equation}
Hence, the finite frequency conductivity in the lab frame and in the comoving frame related by a Galilean transformation are equal.  

For the simplest case, there is no quartic term ($\lambda=0$), so that the system exhibits Galilean relativity. In this case, there is no inversion symmetry breaking in the comoving frame even with a moving CDW state. As a result,  there are no nonreciprocal nonlinear responses. The $k^2$ dispersion is indeed special. In the Method section, we also discuss how the nonlinear transport response can be allowed when the quartic term is further added.


\vspace{1\baselineskip}

\noindent
{\bf The nonreciprocal nonlinear responses in the moving CDW states.}  
Now we are ready to unveil the non-reciprocal nonlinear responses within an intrinsic time-dependent system: the moving CDW states described by the Lagrangian 
$\mathcal{L}_t$
 . As demonstrated earlier, the finite frequency conductivity remains unchanged under the Galilean transformation. Consequently, we can address this challenging problem in the comoving frame using the Lagrangian 
$\mathcal{L}_t'$

To calculate the nonreciprocal nonlinear responses, we first deduce the low-energy Hamiltonian given by $\mathcal{L}_t'$. According to  Eq.~\eqref{main_5} and using $\psi'_{k'_x}=\int dx' e^{-ik_x'x}\psi'(x')$, the energy dispersion in the comoving frame becomes   \begin{equation}
\epsilon_{k'_x}=\frac{k_x'^2}{2m}+4\lambda m^3v^3k_x'+6\lambda m^2v^2k_x'^2+4\lambda mv k_x'^3+\lambda k_x'^4. \label{new2}
\end{equation}

Then we can expand the momentum of the states near the Fermi momentum ($k'_x=\tilde{k}_x+sk_F$ with $s=\pm$ branch) and consider the CDWs would couple these two branches.  The resulting low-energy Hamiltonian (up to the second order in $\tilde{k}_x$) in the comoving frame is given by
\begin{equation}
	H_{0}(\tilde{k}_x)=\begin{pmatrix}
\tilde{\epsilon}_{+,\tilde{k}_x}&\Delta\\
		\Delta& \tilde{\epsilon}_{-,\tilde{k}_x}
\end{pmatrix}\label{Eq_S59}
\end{equation}
where 
\begin{equation}
	\tilde{\epsilon}_{s,\tilde{k}_x}=(s \tilde{v}_F+\delta v_F) \tilde{k}_x+(s\tilde{\beta}+\delta \beta)\tilde{k}_x^2+s\tilde{\alpha} ,
\end{equation}
Here, the coefficients are $\tilde{v}_F=v_F+12\lambda m^2v^2k_F+4\lambda k_F^3, \delta v_F=4\lambda m^3v^3+12\lambda mv k_F^2, 	 \delta \beta=\frac{1}{2m}+6\lambda m^2v^2+6\lambda k_F^2,  \tilde{\alpha}= 4\lambda mv k_F (k_F^2+m^2v^2), \tilde{\beta}=12\lambda mv k_F$. 
Importantly, when $\lambda\neq 0$,    the $s\tilde{\beta} \tilde{k}_x^2$ term that breaks the inversion symmetry is finite, which would be essential for the second-order nonlinear optical responses as we would show next.

\begin{figure}
	\centering
	\includegraphics[width=0.8\linewidth]{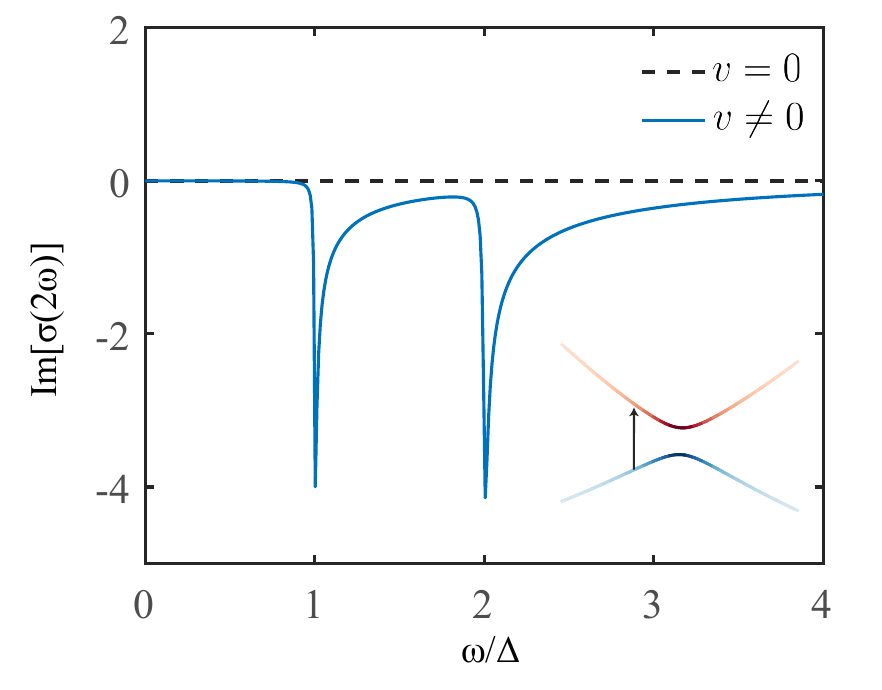}
	\caption{ The second harmonic optical absorptions characterized by $\text{Im}[\sigma(2\omega)]$ (in units of $\frac{\tilde{\beta} e^3}{2\omega^2}$) versus the photon frequency $\omega$, where $\gamma/\Delta=0.01$. The blue and back dashed lines respectively represent the case with CDW motion ($v\neq 0$) and without CDW motion ($v= 0$). The inset schematically shows the optical excitation processes of the CDW folded bands in the comoving frame.}  
\label{fig:fig2}
\end{figure}
Inserting the Hamiltonian $H_0(\tilde{k}_x)$ into the formula of nonlinear optical responses \cite{Moore2019} (see the Method section and Supplementary Note 4 for details), we find that  the second-order nonlinear optical conductivity is given by
\begin{equation}
\sigma^{(2)}(2\omega)=\frac{\tilde{\beta} e^3}{2\omega^2}[2F(\frac{\omega+i\gamma}{\Delta})+F(\frac{2\omega+i\gamma}{\Delta})]. \label{Eq: 12}
\end{equation}
Here, $\omega$ is the photon frequency,     the function $F(a)=\frac{1}{2\pi}[\frac{\pi}{a}+\frac{8}{a\sqrt{a^2-4}}(\text{arctanh}(\frac{a}{\sqrt{a^2-4}})-\text{arctanh}(\frac{a-2}{\sqrt{a^2-4}}))]$, and $F(\frac{\omega+i\gamma}{\Delta})$, $F(\frac{2\omega+i\gamma}{\Delta})$  arise from one-photon and two-photon processes, respectively. The one-photon and two-photon exhibit a leading scaling of $(\omega-2\Delta)^{-1/2}$ and $(\omega-\Delta)^ {-1/2}$, respectively. A typical scaling for photoconductivity in 1D limit. Note that $\gamma$ is a small damping parameter, which is related to the lifetime of an excited electron and can be finite even in a clean system.

The imaginary part of $\sigma(2\omega)$  representing the second harmonic absorption is plotted in Fig.~\ref{fig:fig2}. Note that the real part of $\sigma(2\omega)$ is directly related to $\text{Im}[\sigma(2\omega)]$ according to the Kramers--Kronig relations \cite{Scandolo1995}. The one-photon and two-photon absorption peaks near $\omega=2\Delta$ and $\omega=\Delta$ can be clearly seen.   It is worth noting that $\sigma(2\omega)\propto\tilde{\beta}\propto \lambda mv$, while $\tilde{\beta}$ arises from the cubic term in Eq.~\eqref{new2} induced by the additional quartic term in the lab frame.  Moreover, as expected, the second-order nonreciprocal nonlinear responses are finite only when the CDW states are moving, i.e. $v\neq 0$ so that $\tilde{\beta}\neq 0$. Therefore, we have explicitly demonstrated the nonreciprocal nonlinear responses in the moving CDW states through the Galilean transformation.

\vspace{1\baselineskip}

\noindent
{\bf Discussion} 

\begin{figure}
	\centering
\includegraphics[width=1\linewidth]{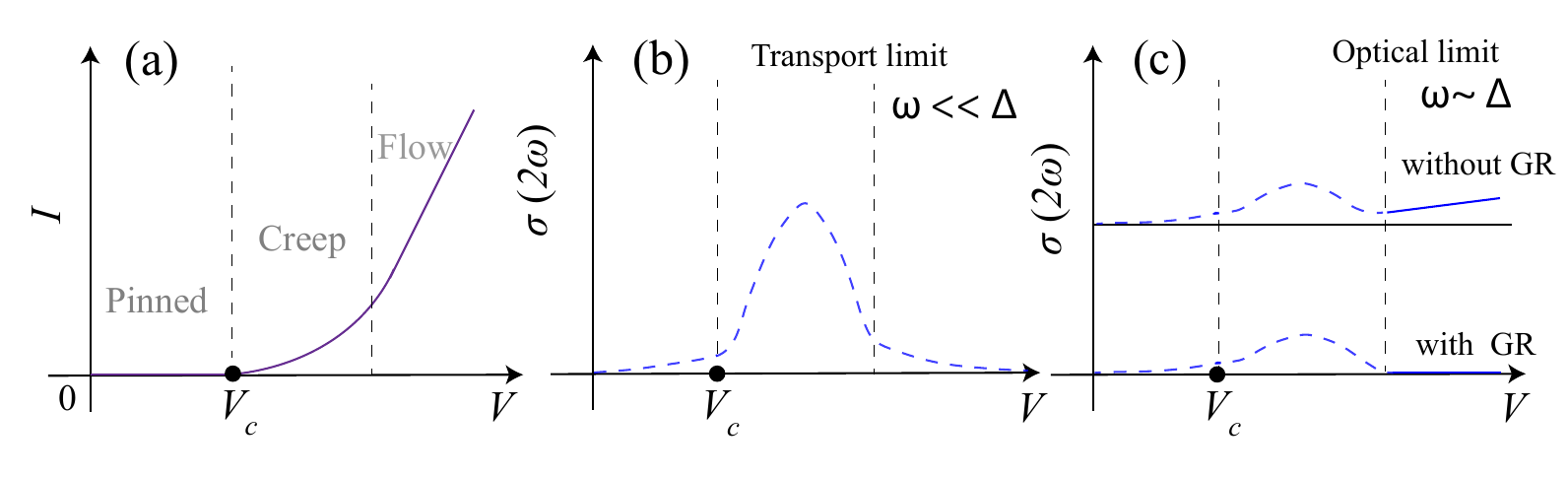}
	\caption{ Schematics of nonlinear responses in moving CDW states. (a) Nonlinear voltage ($V$)- current ($I$) relation to highlight the pinned, creep, and flow region respectively. \textcolor{black}{Here,  $I$ is the DC current induced by CDW motion. The dominant  $I-V$ relation in the flow region is considered to be approximately linear. Note that in general, there may also exhibit some nonlinear corrections due to the presence of impurity pinning effect \cite{Sneddon1982}}.  (b) and (c) schematically depict second-order nonlinear response characterized by $\sigma (2\omega)$ under the electric field in the transport limit ($\omega\ll \Delta$) and the optical limit ($\omega\sim \Delta$). The $\sigma (2\omega)$ of the flow region in the optical limit with and without Galilean relativity (GR) are highlighted as solid blue lines, where our theory applies.    }
	\label{fig:fig3}
\end{figure}

\noindent It is important to note that with an external excitation at optical frequencies above the CDW energy gap, the quasi-particle effects that we have discussed are dominant.  It should be distinguished from the transport region, where the interaction between impurities and collective modes of the CDW should dominate finite frequency responses \cite{Gruner1988,Sneddon1982}.  As shown in Fig.~\ref{fig:fig3}(a), in general, the current-driven CDW states exhibit three distinct regions: the pinned, creep, and flow regions according to the strength of the electric field. The pinned region may exhibit nonlinear responses due to the distortion of the CDW, while the nonlinear responses are expected to peak around the creep region where the depinning motion of CDW is strongest.  Our speculation for the nonlinear responses in the transport region ($\omega\ll \Delta$) is shown in Fig.~\ref{fig:fig3}(b), where the nonlinear responses mostly stem from the creep motion of the sliding density wave. In the optical limit
 ($\omega\sim \Delta$), the quasiparticle effects would be more crucial, which fits our interest in this work. In this case, our results imply that the second-order nonlinear conductivity within the flow region linearly increases with the current in the case without emergent Galilean relativity   [see the solid blue lines in Fig.~\ref{fig:fig3}(c)]. 
 
 We next discuss the some  approximations we have made in our moving CDW model in the main text. First, the continuum model approximation is used in capturing the energy dispersion. We expect that the continuum model approximation is valid when the CDW gap is far away from the Brillouin zone boundary (or much smaller than the bandwidth) so that the lattice periodic potential is not crucial.   Near the Brillouin zone boundary, the underlying periodic lattice potential would become crucial and open the gap in energy dispersion. Second, we have assumed the disorder effects can be negligible in the flow region.  When the Galilean transformation is performed, the presence of disorder potential in the lab frame would result in moving disorders in the comoving frame. However, the CDW order has gapped out the states so that the moving disorders do not have states to scatter near Fermi energy. As a result, we expect that our results based on Galilean transformation would not be affected by the presence of weak disorder effects as long as the speed of moving disorders is not large enough to overcome the CDW gap.

The current-driven  CDW states in the flow region behave as nonequilibrium steady states in this work, which breaks time-reversal and inversion symmetry. We would like to emphasize here that the current carried by the moved CDW  state arises from the time-dependent CDW phase change in real space \cite{Gruner1988}, which resembles the phase gradient-induced current in superconductors.  It should be distinguished from the single-particle currents driven by DC fields in metals. In other words, what is applied externally is the current instead of the DC electric fields for the moving CDW. In practice, due to thermal effects or other dissipations, some small DC electric field may be built in the flow region. The  DC fields can directly couple with the electrons. In insulators, the presence of electric fields can result in the Landau-Zener tunneling \cite{Nagaosa2008}. However, as we estimated in the Supplementary Note 5, the Landau-Zener tunneling probability is negligible for the moving CDW state. Our analysis in the main text thus would not affected by the presence of small DC electric fields with negligible Landau-Zener tunneling effects.     

\textcolor{black}{In the main text, our model is analyzed in the 1D limit. In practice, the system can exhibit a higher dimension. Here, we briefly discuss the 2D limit by introducing the momentum $k_y$. In this case, the normal-state dispersion can be described by the Hamiltonian (in the lab frame): $
H(\bm{k})=\frac{k_x^2}{2m}+\lambda k_x^4-\mu(k_y) 
$, where $\mu(k_y) = \mu + (2t_{\perp} \cos k_y - 2t_{\perp})$, and $t_{\perp}$ is the hopping parameter along the $y$-direction. In this form, the 2D system can be regarded as a set of 1D subbands labeled by $k_y$, with the chemical potential shifted by $(2t_{\perp} \cos k_y - 2t_{\perp})$.  When $4t_{\perp} < \Delta$, the system is expected to remain gapped and behave as quasi-1D. For each $k_y$, the Fermi energy lies within the gap, allowing the system to be treated as a 1D problem. In general, the gap can also exhibit $k_y$ dependence as well. As a result, the inverse square-root singularity would be broadened by $k_y$ integral in Fig.~\ref{fig:fig3}.   On the other hand, in the limit of large hopping ($4t_{\perp} > \Delta$), the energy shift due to hopping along the $y$-direction can overcome the CDW gap, making the system metallic. In this case, the situation becomes more complex, as the presence of low-energy electrons at the Fermi level allows scattering with impurities. As a result, the current would primarily be carried by free electrons rather than the moving CDW. Unfortunately, this scenario is beyond our theoretical consideration.
}

 In summary, we have demonstrated the nonreciprocal nonlinear responses in a nonequilibrium system with time-dependent states---moving CDW states through Galilean transformation. It is important to note that moving CDW states are non-perturbative nonequilibrium steady states. As a result, our analysis is in sharp contrast with the conventional perturbative calculations of non-linear responses in the lab frame.    

 Looking ahead, we encourage experimentalists to revisit the quasi-1D CDW materials such as TaS$_3$, NbSe$_3$, etc. \cite{Gruner1988}, and explore their nonreciprocal nonlinear optical responses above the gap using advanced terahertz light measurements. A comparative analysis of experimental results with our theoretical insights would be intriguing.  Furthermore, we note that the nonreciprocal nonlinear responses would also appear in some other current-driven systems, such as superconductors \cite{Shimano, Liang2022,Papaj2022}, Weyl materials \cite{Moore2021}.  However, it is worth pointing out that there is no clear intrinsic time dependence in the lab frame for these systems, which is in contrast with the moving CDW states. Nevertheless, it would be interesting to revisit the nonlinear nonreciprocal responses in current-driven superconductors or topological materials in the comoving frame with our theoretical framework. One challenge is that our analysis in this work starts from simple models, which may be difficult to apply materials with complicated low-energy dispersions. However, in some materials, the low-energy dispersion can be described with a simple Hamiltonian.  In this case, our theoretical framework based on Galilean transformations can be carried out and offer insights into the external electromagnetic field responses in the nonequilibrium region.          


\noindent{\bf Methods}\\
 \noindent {\bf Galilean transformation on the field operator.} The simplest Lagrangian we are dealing with in this work carries the following form:
 \begin{equation}
\mathcal{L}=i\psi^{\dagger}\partial_t\psi-\frac{1}{2m}\partial_i\psi^{\dagger}\partial_i\psi-\psi^{\dagger}V(x,t)\psi. 
\end{equation}
   Let us perform the Galilean transformation: $x'=x-vt$, $t'=t$, $\frac{\partial}{\partial t}=\frac{\partial}{\partial t'}+\frac{\partial x'}{\partial t}\frac{\partial}{\partial x'}=\frac{\partial}{\partial t'}-v\frac{\partial}{\partial x'}$,  $\frac{\partial}{\partial x}=\frac{\partial}{\partial x'}$. To compensate for the coordinate changes,  the field operator should also been transformed
\begin{eqnarray}
	\psi=e^{i\eta(x',t')}\psi'.
\end{eqnarray}
After the transformation, the Lagrangian becomes
\begin{eqnarray}
	\mathcal{L'}&&=i \psi'^{\dagger}\partial_{t'}\psi'-[\partial_{t'}\eta-v\partial_{x'}\eta+\frac{1}{2m}(\partial_{x'}\eta)^2]\psi'^{\dagger}\psi'\nonumber\\
	&&-iv\psi'^{\dagger}\partial_{x'}\psi'-\frac{i}{2m}(\partial_{x'}\eta)[(\partial_{x'}\psi^{\dagger})\psi'-\psi^{\dagger}\partial_{x'}\psi]\nonumber\\
 &&-\frac{1}{2m} \partial_{x'}\psi'^{\dagger}\partial_{x'}\psi'-\psi'^{\dagger}V'(x',t')\psi'.
\end{eqnarray}
To cancel the  term in $\psi'^{\dagger}\partial_{x'}\psi'$,
\begin{equation}
	-iv+\frac{i}{m}\partial_{x'}\eta=0.
\end{equation}
Then it requires
\begin{equation}
\partial_{x'}\eta=mv
\end{equation}

We can further fix the form of $\eta$ by considering the system is Galilean invariant in the case without spatial dependent potential $V'(x',t')=V(x,t)=0$. In this case, the Lagrangian would exhibit the same form in the lab frame and moving frame if
\begin{equation}
	\partial_{t'}\eta-v\partial_{x'}\eta+\frac{1}{2m}(\partial_{x'}\eta)^2=0.
\end{equation}
Now we obtain the phase factor on the field operator under Galilean transformation is given by
\begin{equation}
	\eta=mvx'+\frac{1}{2}mv^2t'=mvx-\frac{1}{2}mv^2t.
\end{equation}
In the case of moving CDW, the potential $V(x,t)=V(x-vt)$ is mapped to $V(x')$ in the comoving frame after the Galilean transformation.

\noindent{\bf{Nonlinear transport induced by the dispersion without Galilean relativity.}}
 To further highlight that the $k^2$ dispersion is special, we now show the nonlinear transport for the energy dispersion $\epsilon_{k_x}=\frac{k_x^2}{2m}+\lambda k_x^4$ in the absent of the CDW order parameter. We shall consider a DC plus a small AC component of the electric current: $J(t)=J_0+J_{ac}\cos(\omega t)$.  The DC component is to break the inversion symmetry.  It would be expected that the cubic nonlinear responses $V_3=R_3J^3$ ($V_3$ is the voltage, $R_3$ is the resistance) would give rise to a $2\omega$-signal: $V_{2\omega}=3R_3 J_0J_{ac}^2$.

For the moment, we consider the electronic system without CDW but with finite relaxation. From the Boltzmann  transport  equation (see Supplementary Note 3), we can derive that the $2\omega$-component of  
 conductivity given by the third-order nonlinear response is 
\begin{eqnarray}
	\sigma^{(3)}(2\omega)=  \frac{e^4\Gamma(\omega,\tau)  }{4}\int \frac{dk_x}{2\pi}  f^{(0)}\partial_{k_x}^3 v(k_x)=6 e^4 \lambda n \Gamma(\omega,\tau), \nonumber\\\label{main_eq_9}
\end{eqnarray}
where the factor
$
\Gamma(\omega,\tau)= \frac{1}{(2i\omega+\tau^{-1})^2(i\omega+\tau^{-1})}+\frac{\tau}{(2i\omega+\tau^{-1})(i\omega+\tau^{-1})}+\frac{1}{(2i\omega+\tau^{-1})(i\omega+\tau^{-1})^2}$,  $\tau$ is the scattering time, $f^{(0)}$ is the Fermi distribution function, the electron velocity $v(k_x)=\frac{\partial \epsilon(k_x)}{\partial k_x}$, and the electron density $n=\int \frac{d k_x}{2\pi} f^{(0)}(\epsilon(k_x)-\epsilon_F)$. 
From the Eq.~\eqref{main_eq_9}, it can be seen that the nonreciprocal nonlinear responses giving $\sigma(2\omega)$ signal are absent in the simplest quadratic band dispersion ($\lambda=0$). It becomes finite when the quartic term is introduced.

In the main text, we have also obtained the optical second harmonic generation optical conductivity $\sigma(2\omega)$, i.e., Eq.~\eqref{Eq: 12}. Now let us compare the magnitude between these two nonlinear conductivity. It can be seen that 
\begin{equation}
\frac{\sigma^{(2)}(2\omega)}{\sigma^{(3)}(2\omega') E}\sim  \frac{\frac{\tilde{\beta} e^3}{2\omega^2}}{6 e^4\lambda n\Gamma(\omega',\tau) E}
\end{equation}
Here $\omega$ is photon frequency and $\omega'$ is the transport frequency. The numerator only refers to the imaginary part of $\sigma^{(2)}(2\omega)$ here.  Although   there are peaks around $\omega=\Delta$ and $2\Delta$,  $\sigma^{(2)}(2\omega)$ are at order of $\frac{\tilde{\beta} e^3}{2\omega^2}$ (see Fig. 3).
Note that the electric field $E$ that drives the DC current $J_0$ appears on the denominator to fix the dimension. In the optical limit ( $\omega' \gg \tau^{-1})$, $\Gamma(\omega',\tau)\sim \frac{1}{\omega'^3}$. Inserting $\tilde{\beta}=12 mv k_F$, $mv_F=\pi n$, we estimated 
\begin{equation}
\frac{\sigma^{(2)}(2\omega)}{\sigma^{(3)}(2\omega') E} \sim  \frac{\omega'^3}{\omega^2\Delta} \frac{v}{v_F} \frac{\pi \Delta}{ e E k_F^{-1}}.
\end{equation}
Note that $k_F^{-1}\sim a$, where $a$ is the lattice constant. 
Using some reasonable parameters: the CDW moving  velocity $v\sim 100$ m/s,  the Fermi velocity $v_F\sim 10^{6}$ m/s, CDW gap $\Delta \sim 0.1$ eV, the lattice constant $a=3\times 10^{-10}$ m , and the electric field $E= 10^ 5$ V/mm, we expect
\begin{equation}
\frac{\sigma^{(2)}(2\omega)}{\sigma^{(3)}(2\omega') E} \sim  \frac{\omega'^3}{\omega^2\Delta}.
\end{equation}
Therefore, the relative magnitude of these two nonlinear responses is mostly determined by their frequency.
However, it should be noted that these two scenarios are distinct measurements. 

\noindent{\bf Optical responses formalism. }In this Method section, we present the detailed formalism of the linear and second-harmonic optical conductivity. Note that to make the unit of conductance clear, we would keep the $\hbar$ below (which was set to be 1 in the main text).  The linear optical conductivity is given by the bubble diagram \cite{Moore2019},
\begin{eqnarray}
	\sigma^{\mu\alpha}(\omega;\omega)=\frac{i e^2}{\hbar\omega}\sum_{a\neq b}\int  \frac{d^d \bm{k}}{(2\pi)^d} \frac{f_{ab} v_{ab}^{\alpha} v_{ba}^{\mu}}{\omega-\epsilon_{ba}}.
\end{eqnarray} 
where $d$ is the spatial dimension,  the sandwich of velocity operator between interband $v_{ab}^{\alpha}=\braket{a|\partial_{k_\alpha} H(\bm{k})|b}$, the energy and  Fermi distribution difference between the band  $\epsilon_a$ and $\epsilon_b$ are represented as  $\epsilon_{ba}=\epsilon_b-\epsilon_a$, $f_{ab}=f(\epsilon_a)-f(\epsilon_b)$, respectively. 

 The Hamiltonian we are dealing with generally takes the following form:
 \begin{equation}
 	H(\bm{k})=\begin{pmatrix}
 		\xi_{+}(\bm{k}) &\Delta \\
 		\Delta &\xi_{-}(\bm{k})
 	\end{pmatrix}.
 \end{equation} 

 It is easy to show that the eigenvalues of $H'(\bm{k})$ is given by $E_{\pm}=\frac{\xi_{+}(\bm{k})+\xi_{-}(\bm{k})}{2} \pm \epsilon (\bm{k})$ with $\epsilon(\bm{k})=\sqrt{\frac{(\xi_+(\bm{k})-\xi_{-}(\bm{k}))^2}{4}+\Delta^2}$,  the eigenfunctions are $\ket{E_{+}}=(\cos\frac{\theta_{\bm{k}}}{2}, \sin\frac{\theta_{\bm{k}}}{2} )^{T}$, and $\ket{E_{-}}=(-\sin\frac{\theta_{\bm{k}}}{2} ,\cos\frac{\theta_{\bm{k}}}{2})^{T}$ with $\sin \theta_{\bm{k}}=\Delta/\epsilon(\bm{k})$. As a result,  $v^{x}_{+-}(\bm{k})=\braket{E_{+}|\partial_{k_x} H|E_{-}}=\frac{1}{2}\sin\theta_{\bm{k}}(\partial_{k_x}\xi_{-}-\partial_{k_x}\xi_{+})=\frac{\Delta}{2\epsilon(\bm{k})}(\partial_{k_x}\xi_{-}-\partial_{k_x}\xi_{+}).$

The optical conductivity for the second-harmonic generation can be rewritten as \cite{Moore2019}
\begin{eqnarray}
\sigma^{\mu\alpha\beta}(2\omega;\omega,\omega)&&=\sigma_{\text{I}}^{\mu\alpha\beta}(2\omega; \omega,\omega)+\sigma_{\text{II}}^{\mu\alpha\beta}(2\omega; \omega,\omega),\\
	\sigma_\text{I}^{\mu\alpha\beta}&&=-\frac{e^3}{2\hbar^2\omega^2} \sum_{a\neq b}\int [d\bm{k}] f_{ab}\frac{v_{ab}^{\alpha}v_{ba}^{\mu\beta}+v_{ab}^{\beta}v_{ba}^{\mu\alpha}}{\omega-\epsilon_{ab}}\nonumber\\
&&+f_{ab}\frac{v_{ab}^{\alpha\beta}v_{ba}^{\mu}}{2\omega-\epsilon_{ab}}, \label{Eq_S25}\\
	\sigma_{\text{II}}^{\mu\alpha\beta}&&=-\frac{e^3}{2\hbar^2\omega^2} \sum_{a\neq b\neq c}\int [d\bm{k}] \frac{(v_{ab}^{\alpha}v_{bc}^{\beta}+v_{ab}^\beta v_{bc}^{\alpha}) v_{ca}^{\mu}}{\epsilon_{ab}+\epsilon_{cb}} (\nonumber\\&&\frac{2f_{ac}}{2\omega-\epsilon_{ca}}+\frac{f_{cb}}{\omega-\epsilon_{cb}}+\frac{f_{ba}}{\omega-\epsilon_{ba}}) \label{Eq_S26}.
\end{eqnarray}
where $2\omega$ and $\omega$ in Eqs.~\eqref{Eq_S25} and \eqref{Eq_S26} characterize the contributions from two-photon and one-photon processes, respectively. The second-order derivation of Hamiltonian $v_{ba}^{\mu\beta}=\braket{b\bm{k}|\partial_{k_{\mu}}\partial_{k_{\beta}} H(\bm{k})|a\bm{k}}$.  The contribution that involves two (three) different bands is labeled as $\sigma_\text{I}^{\mu\alpha\beta}$ ($	\sigma_\text{II}^{\mu\alpha\beta}$), respectively.  

\noindent {\bf Data availability} 

\noindent All data needed to evaluate the conclusions in the paper are present in the paper.

\noindent {\bf Acknowledgements}

\noindent  
N.N. was supported by JSPS KAKENHI Grant No. 24H00197 and 24H02231.
N.N. was also supported by the RIKEN TRIP initiative.  Y.M.X.  acknowledges financial support from the RIKEN Special Postdoctoral Researcher(SPDR) Program. 

\noindent {\bf Author contributions}

\noindent N.N. initiated and guided this work. H.I. helped to analyze the problem. Y.M.X. carried out the calculations and wrote the manuscript with suggestions from all the authors.

\noindent {\bf Competing interests}

\noindent The authors declare no competing interests.

 \end{document}